\documentclass[12pt]{article}
\usepackage{epsf}
\newcommand{\be}{\begin{equation}}
\newcommand{\ee}{\end{equation}}
\newcommand{\bea}{\begin{eqnarray}}
\newcommand{\eea}{\end{eqnarray}}
\newcommand{\bfg}{\begin{figure}[htbp]}
\newcommand{\efg}{\end{figure}}
\newcommand{\lb}{\label}
\newcommand{\rf}{\ref}

\begin{document}

\begin{flushright}
IPNO/DR 00-08
\end{flushright}

\begin{center}
{\large \textbf{The pionium lifetime\\
in generalized chiral perturbation theory}}    
\vspace{1 cm}

H. Sazdjian \\
\textit{Groupe de Physique Th\'eorique, Institut de Physique Nucl\'eaire,\\
Universit\'e Paris XI, F-91406 Orsay Cedex, France\\
E-mail: sazdjian@ipno.in2p3.fr}
\end{center}
\vspace{1 cm}

\begin{center}
{\large Abstract}
\end{center}

Pionium lifetime corrections to the nonrelativistic formula are calculated
in the framework of the quasipotential--constraint theory approach.
The calculation extends an earlier evaluation, made in the scheme of standard
chiral perturbation theory, to the scheme of generalized chiral
perturbation theory, in which the quark condensate is left as a free
parameter. The pionium lifetime is calculated as a function of the
combination $(a_0^0-a_0^2)$ of the $\pi\pi$ $S$-wave scattering lengths
with isospin $I=0,\ 2$. 
\par

\vspace{1.5 cm}

The DIRAC experiment at CERN \cite{dir} is expected to allow, in the near
future, the measurement of the lifetime of the pionium ($\pi^+\pi^-$ atom)
with a 10\% precision. The latter, in turn, through its relationship
with $\pi\pi$ scattering lengths, would provide a determination
of the combination $(a_0^0-a_0^2)$ of the $S$-wave scattering lengths
with isospin $I=0,\ 2$ with 5\% accuracy. The strong interaction scattering
lengths $a_0^0$ and $a_0^2$ have been evaluated in the literature in the
framework of chiral perturbation theory ($\chi PT$) to two-loop order of the
chiral effective lagrangian \cite{gl,bcegs,kmsf}. Therefore, the pionium 
lifetime measurement provides a high precision experimental test of chiral 
perturbation theory predictions.
\par
The nonrelativistic formula of the pionium lifetime in lowest order of
electromagnetic interactions was first evaluated by Deser \textit{et al.}
\cite{dgbt} and later reanalyzed by others \cite{uptbnnteilbpt}. It
reads:
\be \lb{e1}
\frac{1}{\tau_0} = \Gamma_0 = \frac{16\pi}{9}\sqrt
{\frac{2\Delta m_{\pi}}
{m_{\pi^+}}} \frac{(a_0^0-a_0^2)^2}{m_{\pi^+}^2} |\psi_{+-}(0)|^2,\ \ \
\ \ \Delta m_{\pi}=m_{\pi^+}-m_{\pi^0},
\ee
where $\psi_{+-}(0)$ is the wave function of the pionium at the origin
(in $x$-space).
\par
An evaluation of the relativistic and higher-order electromagnetic 
corrections to the above formula was recently done by several authors.
In the frameworks of quantum field theory and $\chi PT$, three different
methods of evaluation have led to similar estimates, of the order of 6\%, of
these corrections \cite{jss,illr,gglr}. The first method uses a 
three-dimensionally reduced form of the Bethe--Salpeter equation 
(the quasipotential--constraint theory approach) and deals 
with an off-mass shell formalism \cite{jss}. The second method uses the
Bethe--Salpeter equation with the Coulomb gauge \cite{illr}. The third one
uses the approach of nonrelativistic effective theory 
\cite{gglr,cl,lb,es,aggr}. (A survey of other approaches is presented in
the second paper of Ref. \cite{gglr}.)
\par
The pionium lifetime, with the sizable relativistic and electromagnetic
corrections included in, can be represented as:
\bea \lb{e2}
\frac{1}{\tau}=\Gamma&=&\frac{1}{64\pi m_{\pi^+}^2}
\big(\mathcal{R}e\widetilde{\mathcal{M}}_{00,+-}\big)^2 (1+\gamma)
|\psi_{+-}(0)|^2 \sqrt{\frac
{2\Delta m_{\pi}}{m_{\pi^+}}(1-\frac{\Delta m_{\pi}}{2m_{\pi^+}})}
\nonumber \\
&\equiv& \Gamma_0\sqrt{(1-\frac{\Delta m_{\pi}}{2m_{\pi^+}})}
\Big(1+\frac{\Delta \Gamma}{\Gamma_0}\Big),
\eea
where $\mathcal{R}e\widetilde{\mathcal{M}}_{00,+-}$ is the real part of
the on-mass shell scattering amplitude of the process $\pi^+\pi^-\rightarrow
\pi^0\pi^0$, calculated at threshold, in the presence of electromagnetic
interactions and from which singularities of the infra-red photons have
been appropriately subtracted \cite{ku}; the factor $\gamma$ represents the
contribution of interactions at second-order of perturbation
theory with respect to the nonrelativistic zeroth-order Coulomb
hamiltonian. The explicit expressions of
$\mathcal{R}e\widetilde{\mathcal{M}}_{00,+-}$ and of $\gamma$ may differ
from one approach to the other, according to the way the singularities
of the infra-red photons are subtracted, but their total contribution should 
be the same.
\par
The purpose of the present article is to calculate the pionium lifetime
in the framework of generalized $\chi PT$ \cite{fss,kmsf}. In this
framework, according to the observation that the fundamental order
parameter of spontaneous chiral symmetry breaking in QCD is $F_{\pi}$, the
decay coupling constant of the pion, the quark condensate
in the chiral limit $<0|\overline qq|0>_0$ is left as a free parameter.
Its value should  depend on the details of the mechanism of chiral
symmetry breaking and would require an independent experimental test. 
In the framework of standard $\chi PT$, the mechanism of chiral symmetry
breaking would be very similar to that realized in a ferromagnetic medium
\cite{l}. In this case, the value of the Gell-Mann--Oakes--Renner (GOR) 
parameter \cite{gmorgw}, defined as
\be \lb{e3}
x_{GOR}=-\frac{2\hat m<0|\overline qq|0>_0}{F_{\pi}^2m_{\pi}^2},
\ee
where $2\hat m=m_u+m_d$ and $m_{\pi}$ is the pion physical mass, would be
close to one. Equivalently, the quark condensate parameter 
\be \lb{e4}
B\equiv -\frac{<0|\overline qq|0>_0}{F^2},
\ee
where $F$ is $F_{\pi}$ in the chiral $SU(2)\times SU(2)$ limit, 
would be of the order of the hadronic mass scale $\Lambda_H\sim 1$ GeV. This
assumption fixes the way standard $\chi PT$ is expanded: the quark
condensate parameter $B$ is assigned dimension zero in the infra-red 
external momenta of the Goldstone bosons, while the quark masses are 
assigned dimension two \cite{gl}.
\par
On the other hand, in an antiferromagnetic medium, one meets a situation
where the quark condensate would be zero or very small \cite{l,s}. Recently
other possibilities were also advocated \cite{m,dgs}: the existence of a 
possible chiral phase transition in QCD \cite{bz,atwsvsgg} at relatively 
low values of the light quark flavor number might induce, in the standard 
case, a strong flavor dependence of the quark condensate, which would have 
the tendancy to decrease by passing from $SU(2)\times SU(2)$ to 
$SU(3)\times SU(3)$.
\par
The framework of generalized $\chi PT$ offers the possibility of
experimentally testing various issues of chiral symmetry breaking mechanism.
In this framework one relaxes the GOR assumption and treats the
order of magnitude of the quark condensate parameter $B$ as an
\textit{a priori} unknown quantity (awaiting a precise experimental
information about it) leaving to it the possibility of reaching small
or vanishing values. To this aim, $B$ is assigned dimension one in the
infra-red momenta of the external Goldstone bosons and accordingly quark
masses are also assigned dimension one. Due to this rule, at each order
of the perturbative expansion, generalized $\chi PT$ contains more terms
than standard $\chi PT$. For instance, the pion mass formula becomes at
leading order:
\be \lb{e5}
(m_{\pi}^2)_0=2\hat mB+4\hat m^2A,
\ee
where the constant $A$ is expressible in terms of two-point functions
of scalar and pseudoscalar quark densities. In the standard $\chi PT$ case,
this term is relegated to the next-to-leading order.
\par
Processes involving only Goldstone bosons are sensitive at leading order
to the value of the quark condensate parameter $B$. Among them, the
$\pi\pi$ scattering amplitude plays a key role. At the tree level of the
chiral effective lagrangian, the amplitude $A(s|t,u)$ has the expression:
\be \lb{e6}
A(s|t,u)=\frac{1}{F^2}(s-2\hat mB),
\ee
which displays explicit dependence on the quark condensate parameter.
\par
At the one-loop order the strong interaction on-mass shell scattering 
amplitude can be described by four parameters, $\alpha$, $\beta$, $\lambda_1$
and $\lambda_2$, and takes the following form \cite{kmsf,g}:
\bea \lb{e7}
A(s|t,u)&=&\mathcal{M}_{00,+-}^{str.}=
\frac{\beta}{F_{\pi}^2}(s-\frac{4}{3}m_{\pi}^2)+\alpha
\frac{m_{\pi}^2}{3F_{\pi}^2}\nonumber \\
&+&\frac{\lambda_1}{F_{\pi}^4}(s-2m_{\pi}^2)^2+\frac{\lambda_2}{F_{\pi}^4}
\Big[(t-2m_{\pi}^2)^2+(u-2m_{\pi}^2)^2\Big]\nonumber \\
&+&\frac{1}{2F_{\pi}^4}\Big[s^2-m_{\pi}^4+8(4\hat m^2A)s-6(4\hat m^2A)
m_{\pi}^2+7(4\hat m^2A)^2\Big]\bar J(s)\nonumber \\
&+&\frac{1}{F_{\pi}^4}\Big[(m_{\pi}^2+4\hat m^2A-\frac{1}{2}t)^2+
\frac{1}{12}(s-u)(t-4m_{\pi}^2)\Big]\bar J(t)\nonumber \\
&+&\frac{1}{F_{\pi}^4}\Big[(m_{\pi}^2+4\hat m^2A-\frac{1}{2}u)^2+
\frac{1}{12}(s-t)(u-4m_{\pi}^2)\Big]\bar J(u), 
\eea
where $m_{\pi}$ is the physical pion mass, $\bar J$ the conventional
loop function \cite{gl} and $4\hat m^2A$ the quadratic mass term
present at the tree level in $m_{\pi}^2$ [Eq. (\rf{e5})]. The parameters
$\lambda_1$ and $\lambda_2$ are related to the standard renormalized
low energy constants $l_1^r$ and $l_2^r$ \cite{gl}; their expressions 
(at the one-loop order) are:
\bea
\lb{e8}
\lambda_1&=&2l_1^r-\frac{1}{48\pi^2}\ln(\frac{m_{\pi}^2}{\mu^2})-
\frac{1}{36\pi^2}, \\
\lb{e9}
\lambda_2&=&l_2^r-\frac{1}{48\pi^2}\ln(\frac{m_{\pi}^2}{\mu^2})-
\frac{5}{288\pi^2}.
\eea
($\mu$ is the renormalization mass.) $\beta$ is essentially related to the 
deviation of $F_{\pi}$ from $F$. $\alpha$ is mainly related
to the quark condensate parameter;  its expression at the tree level is:
\be \lb{e10}
(\alpha)_0=1+3\times \frac{4\hat m^2A}{(m_{\pi}^2)_0}.
\ee
The one-loop expressions of $F_{\pi}^2$, $m_{\pi}^2$, $\beta$ and $\alpha$
are given by the following relations:
\bea 
\lb{e10a}
F_{\pi}^2&=&F^2\Big[1+2l_4^r\frac{2\hat mB}{F^2}+j_1^r
\frac{4\hat m^2A}{F^2}-\frac{2m_{\pi}^2}{16\pi^2}\ln(\frac{m_{\pi}^2}
{\mu^2})\Big],\\
\lb{e10b}
m_{\pi}^2&=&2\hat mB+4\hat m^2A+2l_3^r
\frac{(2\hat mB)^2}{F^2}\nonumber \\
& &+j_2^r(2\hat mB)\frac{4\hat m^2A}{F^2}+j_3^r\frac{(4\hat m^2A)^2}{F^2}
\nonumber \\
& &-\frac{2}{16\pi^2F^2}(2\hat mB+4\hat m^2A)\Big(-\frac{1}{4}(2\hat mB)+
4\hat m^2A\Big)\ln(\frac{m_{\pi}^2}{\mu^2}),\\
\lb{e10c}
\frac{\beta}{F_{\pi}^2}&=&\frac{1}{F^2}\Big[1-\frac{2m_{\pi}^2}
{16\pi^2F^2}-\frac{4\hat m^2A}{F^2}\Big(j_4^r+\frac{5}{16\pi^2}
\big(1+\ln(\frac{m_{\pi}^2}{\mu^2})\big)\Big)\Big],\\
\lb{e10d}
\alpha\frac{m_{\pi}^2}{F_{\pi}^2}&=&
\frac{1}{F^2}\Big(2\hat mB+4(4\hat m^2A)\Big)
+8l_3^r\frac{(2\hat mB)^2}{F^4} \nonumber \\
& &+j_5^r\frac{1}{F^4}(2\hat mB)(4\hat m^2A)+j_6^r\frac{(4\hat m^2A)^2}
{F^4}\nonumber \\
& &-\frac{1}{16\pi^2F^4}\Big[-2m_{\pi}^4+\frac{27}{2}m_{\pi}^2(4\hat m^2A)
+\frac{33}{2}(4\hat m^2A)^2\Big]\ln(\frac{m_{\pi}^2}{\mu^2})\nonumber \\
& &-\frac{1}{16\pi^2F^4}\Big[\frac{1}{2}m_{\pi}^4+11m_{\pi}^2(4\hat m^2A)
+\frac{33}{2}(4\hat m^2A)^2\Big],
\eea
where $j_i^r$ are combinations of renormalized low energy constants
present in the generalized $\chi PT$ lagrangian \cite{kmsf,g,akkt} and
$l_3^r$ and $l_4^r$ are the renormalized low energy constants already present
in the standard $\chi PT$ lagrangian \cite{gl}.
[Our effective lagrangian slightly differs from that of Refs.
\cite{g,akkt} in that we continue using in it the $l_4$-term, instead of
replacing it through the equations of motion with the $\xi^{(2)}$-term.]
\par
In terms of the above parameters the expressions of the $S$- and $P$-wave 
scattering lengths are:
\bea
\lb{e11}
& &a_0^0=\frac{1}{96\pi}\frac{m_{\pi}^2}{F_{\pi}^2}(5\alpha+16\beta)
+\frac{5}{8\pi}\frac{m_{\pi}^4}{F_{\pi}^4}(\lambda_1+2\lambda_2)
+\frac{1}{4608\pi^3}\frac{m_{\pi}^4}{F_{\pi}^4}(5\alpha+16\beta)^2,\\
\lb{e12}
& &a_0^2=\frac{1}{48\pi}\frac{m_{\pi}^2}{F_{\pi}^2}(\alpha-4\beta)
+\frac{1}{4\pi}\frac{m_{\pi}^4}{F_{\pi}^4}(\lambda_1+2\lambda_2)
+\frac{1}{1152\pi^3}\frac{m_{\pi}^4}{F_{\pi}^4}(\alpha-4\beta)^2,\\
\lb{e13}
& &a_1^1=\frac{1}{24\pi}\frac{1}{F_{\pi}^2}\beta-\frac{1}{6\pi}
\frac{m_{\pi}^2}{F_{\pi}^4}(\lambda_1-\lambda_2)+\frac{1}{41472\pi^3}
\frac{m_{\pi}^2}{F_{\pi}^4}(5\alpha^2-40\alpha\beta-16\beta^2).
\eea
\par
At the two-loop order, the $\pi\pi$ scattering amplitude is described by six
parameters, $\alpha$, $\beta$, $\lambda_1$, $\lambda_2$, $\lambda_3$ and
$\lambda_4$. The five parameters other than $\alpha$ are weakly dependent
on the quark condensate; thus $\alpha$ remains the only parameter to be
strongly sensitive to the quark condensate value. It is evident from the
tree level relation (\rf{e10}) that in the standard case $\alpha$ remains
close to 1, while in the extreme case of generalized $\chi PT$
(antiferromagnetic case), where the quark condensate vanishes
($2\hat mB\simeq 0$), it approaches the value 4. The expressions 
of the scattering lengths and of the effective ranges at the two-loop
order in terms of the above six parameters can be found in Ref. \cite{kmsf}.
\par
In order to obtain the pionium lifetime, one first needs to calculate
$\mathcal{R}e\widetilde\mathcal{M}_{00,+-}$ [Eq. (\rf{e2})]. This includes,
in addition to the strong interaction contributions, those of the
electromagnetic interactions. These effects are taken into account in the
chiral effective lagrangian by the presence of new terms involving low
energy constants \cite{u,mms,ku}. In generalized $\chi PT$ the effective
lagrangian contains more terms than the standard one, these being proportional
to the low energy parameter $A$ [Eqs. (\rf{e5}) and (\rf{e10})]. The additional 
elctromagnetic terms needed for the present process are given, in standard
notations, by the following part of the effective lagrangian (in the
$SU(2)\times SU(2)$ case):
\bea \lb{e14}
\mathcal{L}^{(G\chi PT)}&=&\frac{1}{2}F^2A\bigg\{k_{15}\langle QUQU^{\dagger}
\rangle  \langle \chi^{\dagger}U+U^{\dagger}\chi\rangle^2\nonumber \\
& &\ \ \ \ \ \ \ \ \ \ +8k_{16}\langle QUQU^{\dagger}\rangle 
\langle\chi^{\dagger}\chi\rangle+k_{17}\langle Q^2\rangle 
\langle \chi^{\dagger}U+U^{\dagger}\chi\rangle^2 \nonumber\\
& &\ \ \ \ \ \ \ \ \ \ +k_{19}\langle \Big(QU\chi^{\dagger}U+U^{\dagger}
\chi U^{\dagger}Q\Big)^2\rangle +k_{20}\langle QU\chi^{\dagger}U\rangle 
\langle U^{\dagger}\chi U^{\dagger}Q\rangle \bigg\},
\eea
where $Q$ is the quark charge matrix, $Q=e\times$diag.$(2/3,-1/3)$,
and the sources $\chi$ have been restricted to a combination of an isosinglet
scalar density and an isotriplet of pseudoscalar densities,
$\chi=s+i\mathbf{p}.\mbox{\boldmath$\tau$}$, with $s$ and $\mathbf{p}$
real, \mbox{\boldmath$\tau$} representing the Pauli matrices. The
renormalization of the coefficients $k$ is done, with dimensional
regularization in $d$-dimensional space-time, according to the decomposition
\be \lb{e15}
k_i=\kappa_i\lambda+k_i^r(\mu),\ \ \ \ \ \ 
\lambda=\frac{\mu^{d-4}}{16\pi^2}\Big(\frac{1}{d-4}-\frac{1}{2}(\ln(4\pi)
+\Gamma'(1)+1)\Big),
\ee
$\mu$ being the renormalization mass. The $\mu$-dependence of the
renormalized coefficients $k_i^r(\mu)$ is fixed by the prescription that
the $k$s are $\mu$-independent. The coefficients $\kappa$ are found to
have the following values:
$\kappa_{15}=\frac{1}{2}+6Z$, $\kappa_{16}=-Z$,
$\kappa_{17}=-\frac{1}{2}-\frac{12}{5}Z$,
$\kappa_{19}=2$, $\kappa_{20}=5$,
where $Z=C/F^4$, $C$ being the coefficient of the lowest-order, $O(e^2p^0)$,
electromagnetic term 
\be \lb{e17}
\mathcal{L}^{(e^2p^0)}=C\langle QUQU^{\dagger}\rangle,
\ee
responsible for the pion mass shift at that order \cite{dgmly}:
\be \lb{e18}
(\Delta m_{\pi}^2)_0=2e^2\frac{C}{F^2},\ \ \ \ \ \ \Delta m_{\pi}^2
\equiv m_{\pi^+}^2-m_{\pi^0}^2.
\ee
We use in the following for the other coefficients $k$ the notations of Ref.
\cite{mms}.
\par
Treating the electromagnetic interaction in
$\mathcal{R}e \widetilde \mathcal{M}_{00,+-}$ as a perturbation yields for
the zeroth-order part of the latter the strong interaction
amplitude, which is already calculated in generalized $\chi PT$ to 
two-loop order \cite{kmsf}. It is therefore sufficient to calculate,
up to one-loop order of the chiral effective lagrangian,
the first-order electromagnetic correction to it. It should be emphasized
here that, at the numerical level, the strong interaction amplitude is 
calculated in the literature \cite{gl,bcegs,kmsf} by identifying the pion
mass appearing in it with the physical mass of the charged pion.
\par
The electromagnetic corrections can be divided into two categories. The
first one concerns corrections that arise from conventional pion-photon
interactions. The second one concerns corrections that arise from
quark-photon interactions which manifest themselves in the chiral effective
lagrangian through the presence of the $O(e^2p^0)$ mass shift term of the 
charged pion [Eq. (\rf{e17})], which survives the chiral limit and produces
the main part of the pion mass difference \cite{dgmly}. (It also induces,
through renormalization, higher-order counterterms in the chiral effective
lagrangian.)
\par
The pion-photon interaction yields in general an infra-red divergent
amplitude on the mass-shell. Usually, this divergence is avoided by giving
the photon a small mass and then subtracting the singular pieces \cite{ku}.
In our approach, based on the quasipotential--constraint theory method
\cite{jss}, we use an off-maas shell formalism, where the total energy of 
the two-pion system is fixed at the bound state energy, that is, below 
thereshold. In this case the photon field can be taken massless and the
corresponding scattering amplitude is finite. However, it still contains a
spurious finite infra-red term (with respect to the bound state problem).
This term is cancelled by the presence of three-dimensional diagrams, 
called constarint diagrams, the role of which is to trasnform, through the
Lippmann--Schwinger equation, the scattering amplitude into an irreducible 
kernel or a potential. The sum of the two contributions is finite and free 
of spurious terms \cite{jss}. It then can be continued, neglecting $O(e^4)$
terms, to the threshold of the $\pi^+\pi^-$ system.
\par
The quark-photon interaction terms are free of infra-red singularities
and the corresponding part of the scattering amplitude can be readily
calculated, taking into account in particular the pion mass difference.
\par
The electromagnetic corrections to the neutral and charged pion masses
are given to order $e^2p^2$ by the following formulas:
\bea 
\lb{e19}
m_{\pi^0}^2&=&m_{\pi}^2
+2e^2\frac{C}{16\pi^2F^4}(2\hat mB)(1+\ln(\frac{m_{\pi}^2}{\mu^2}))
\nonumber \\
& &+e^2(2\hat mB)\Big(-\frac{20}{9}(k_2^r+k_{10}^r)+2(2k_3^r+k_4^r)
+\frac{20}{9}(k_7^r+k_{11}^r)
\Big)\nonumber \\
& &+e^2(4\hat m^2 A)\Big(-\frac{20}{9}(k_2^r+k_{10}^r)+
2(2k_3^r+k_4^r)+\frac{20}{9}(k_{15}^r+k_{17}^r+k_{19}^r)-\frac{8}{9}
k_{20}^r\Big),\nonumber \\
\\
\lb{e20}
m_{\pi^+}^2&=&m_{\pi^0}^2+2e^2\frac{C}{F^2}+e^2\frac{m_{\pi}^2}
{16\pi^2}\big(7-3\ln(\frac{m_{\pi}^2}{\mu^2})\big)\nonumber \\
& &-2e^2\frac{C}{16\pi^2F^4}\Big[(2\hat mB)(1+3\ln(\frac{m_{\pi}^2}{\mu^2}))
+(4\hat m^2A)(4+6\ln(\frac{m_{\pi}^2}{\mu^2}))\Big]\nonumber \\
& &+e^2(2\hat mB)\Big(-2(2k_3^r+k_4^r)+2(k_7^r-2k_8^r)+6(k_7^r+2k_8^r)
\Big) \nonumber \\
& &+e^2(4\hat m^2A)\Big(-2(2k_3^r+k_4^r)+4k_{15}^r+4k_{16}^r
-2k_{19}^r+k_{20}^r\Big),
\eea
where $m_{\pi}$ is the strong interaction mass (\rf{e10b}).
\par
We present in the following the difference,
$\Delta \mathcal{R}e\mathcal{M}_{00,+-}$, at the $\pi^+\pi^-$ threshold,
of $\mathcal{R}e\widetilde\mathcal{M}_{00,+-}$ [Eq. (\rf{e2})] calculated
in the presence of electromagnetism including the terms of order $e^2p^0$
and $e^2p^2$ and of the strong interaction amplitude 
$\mathcal{R}e\mathcal{M}_{00,+-}^{str.}$ [Eq. (\rf{e7})] calculated with the
physical charged pion mass:
\bea \lb{e21}
\Delta \mathcal{R}e\mathcal{M}_{00,+-}&=&4\big(\frac{F_{\pi}^2}{F^2}-
\beta\big)\frac{\Delta m_{\pi}^2}{F_{\pi}^2}+\Big(2(4\beta-\alpha)-
(4-(\alpha)_0)\Big)\frac{\Delta m_{\pi}^2}{3F_{\pi}^2}\nonumber \\
& &+\frac{m_{\pi}^2\Delta m_{\pi}^2}{16\pi^2F_{\pi}^4}
\Big[\Big(\frac{4\beta-\alpha}{3}\Big)^2\big(1+\ln(\frac{m_{\pi}^2}
{\mu^2})\big)-\frac{2\beta}{3}(8\beta+\alpha)\ln(\frac{m_{\pi}^2}{\mu^2})
\Big]\nonumber \\
& &-e^2\frac{m_{\pi}^2}{48\pi^2F_{\pi}^2}(8\beta+\alpha)\big(5+3\ln(\frac
{m_{\pi}^2}{\mu^2})\big)\nonumber \\
& &+\frac{m_{\pi}^2\Delta m_{\pi}^2}{F_{\pi}^4}
\beta\Big[4\beta\lambda_1-\frac{2}{3}(\alpha-\beta)(\frac{1}{2}a_2^r+b_1^r)
\Big]\nonumber \\
& &+\frac{m_{\pi}^2\Delta m_{\pi}^2}{16\pi^2F_{\pi}^4}\big[9\beta^2
-8\alpha\beta-2\alpha^2-\frac{1}{27}(7\beta^2+10\alpha\beta+7\alpha^2)\big]
\nonumber \\
& &+2\beta e^2\frac{m_{\pi}^2}{F_{\pi}^2}\Big\{\big[\frac{1}{3}
(2k_2^r-10k_{10}^r)-2(2k_3^r+k_4^r)+(k_7^r-2k_8^r)+3(k_7+2k_8^r)\big]
\nonumber \\
& &\ \ \ \ \ \ +\frac{(\alpha-\beta)}{3\beta}\big[\frac{10}{9}
(k_2^r+k_{10}^r)-\frac{10}{9}(k_7^r+k_{11}^r)-(k_7^r-2k_8^r)\nonumber \\
& &\ \ \ \ \ \ \ \ \ \ -3(k_7^r+2k_8^r)-\frac{1}{9}(46k_{15}^r+10k_{17}^r
+32k_{19}^r-11k_{20}^r)\big]\nonumber \\
& &\ \ \ \ \ \ +\frac{2}{9}\Big(\frac{\alpha-\beta}{3\beta}\Big)^2
\big[5(k_7^r+k_{11}^r)-5(k_{15}^r+k_{17}^r+k_{19}^r)+2k_{20}^r\big]
\Big\}.
\eea
Here $(\alpha)_0$ represents $\alpha$ in the chiral $SU(2)\times SU(2)$
limit [Eq. (\rf{e10})]; $m_{\pi}$ and $F_{\pi}$ correspond to the strong
interaction quantities [Eqs. (\rf{e10a})-(\rf{e10b})]; $a_2^r$ and $b_1^r$
are low energy constants appearing in the generalized $\chi PT$ lagrangian
\cite{g,akkt}; $\Delta m_{\pi}^2=m_{\pi^+}^2-m_{\pi^0}^2$.
\par
We list below the remaining corrections to the pionium decay width 
\cite{jss}. (We designate by $\alpha^{em}$ the fine structure constant in
order to distinguish it from the parameter $\alpha$ introduced previously.)
\par
1) The strong interaction correction coming from second order perturbation
theory with respect to the bound state wave equation is:
\be \lb{e22}
\big(\Delta \Gamma\big)_{str.}=1.5\alpha^{em}(2a_0^0+a_0^2)\Gamma_0.
\ee
\par
2) The vacuum polarization correction is:
\be \lb{e23}
\big(\Delta \Gamma\big)_{vac.\ pol.}=0.41\alpha^{em}\Gamma_0.
\ee
\par
3) An effective $O(e^2p^2)$ correction also arises from formal $O(e^2p^4)$
effects, corresponding to diagrams with one pion loop and one photon 
propagator. In the present bound state formalism, the diagram with one pion
loop with one photon exchange provides an infra-red logarithmic contribution,
which is:
\bea \lb{e24}
\big(\Delta \Gamma\big)_{O(e^2p^4)}&=&-\frac{\alpha^{em}}{3}(2a_0^0+a_0^2)
\Big(2\ln\alpha^{em}+3\ln 2+21\zeta(3)/(2\pi^2)\Big)\Gamma_0\nonumber \\
&=&2.2\alpha^{em}(2a_0^0+a_0^2)\Gamma_0.
\eea
This effect could be considered as being part of
$\mathcal{R}e\widetilde\mathcal{M}_{00,+-}$ [Eq. (\rf{e2})]; however, since
$\Delta \mathcal{R}e\mathcal{M}_{00,+-}$ in Eq. (\rf{e21}) has been defined
as including only one-loop graphs, we write it here separately.
\par
4) The corrections due to isospin breaking in the quark masses being
quadratic in $(m_u-m_d)$ \cite{gl}, provide negligible effects and are
ignored.
\par
5) The kinematic correction coming from the phase space factor 
[Eq. (\rf{e2})] is not included in the definition of 
$\Delta \Gamma/\Gamma_0$ and should be directly incorporated in
$\Gamma$ or $\tau$. 
\par
Collecting all these contributions, one finds for the total dynamical
correction to the nonrelativistic decay width formula 
[Eqs. (\rf{e1})-(\rf{e2})] the following expression:
\be \lb{e25}
\frac{\Delta \Gamma}{\Gamma_0}=
\frac{6\Delta \mathcal{R}e\mathcal{M}_{00,+-}}{32\pi(a_0^0-a_0^2)}
+3.7\alpha^{em}(2a_0^0+a_0^2)+0.41\alpha^{em},
\ee
where $a_0^0$ and $a_0^2$ are the strong interaction scattering lengths
calculated up to two loops in the framework of generalized $\chi PT$
\cite{kmsf}.
\par
In order to evaluate the pionium lifetime, one first needs to know the
value of the combination $(a_0^0-a_0^2)$ of the scattering lengths, which
enters in the nonrelativistic formula (\rf{e1}) as well as in the correction
(\rf{e25}). The evaluation of the latter necessitates also the knowledge
of the other parameters of the theory. Among the six parameters ($\alpha$,
$\beta$, $\lambda_i$, $i=1,\ldots,4$) of the strong interaction amplitude
five of them ($\beta$ and the $\lambda$s) are weakly dependent on the
quark condensate value. It is then natural to fix them at some mean values
and to consider $\alpha$ as the only variable in the preceding relations.
The precise values of the six parameters depend on the behavior of the
$\pi-\pi$ scattering amplitude in the whole low energy kinematic region.
A determination of these values in the standard $\chi PT$ case was
presented in Ref. \cite{gkms}. In the present problem, however, the
main quantity of interest is the combination $(a_0^0-a_0^2)$ of the
scattering lengths, rather than the general scattering amplitude itself.
Our aim is to establish a relationship between the pionium lifetime and
this combination of the scattering lengths which could serve us to extract
the value of $(a_0^0-a_0^2)$, within an uncertainty interval, from the
experimental value of the pionium lifetime. Furthermore, we want to place
the prediction of standard $\chi PT$ on the central line of the above
relationship.
\par
To achieve this, we have fixed the values of the parameters $\beta$ and
$\lambda_i$ ($i=1,\ldots,4$) from the predictions of standard $\chi PT$
\cite{bcegs}. We have considered set I of threshold parameters, for which
one has in particular $(a_0^0-a_0^2)=0.258$, and determined the
parameters from the values of $a_0^0$, $a_0^2$, $a_1^1$, $a_2^0$, $b_0^2$
and $b_1^1$. We have obtained the following values for the parameters:
$\alpha=1.021$, $\beta=1.109$, $\lambda_1=-10.28\times 10^{-3}$,
$\lambda_2=15.90\times 10^{-3}$, $\lambda_3=0.81\times 10^{-4}$,
$\lambda_4=-1.00\times 10^{-4}$. These values should be considered as a
convenient means of analyzing the present problem, rather than the best
values for the entire scattering amplitude; their role is to introduce
the standard $\chi PT$ prediction as an initial data placed on the central
line of the relationship between the lifetime and $(a_0^0-a_0^2)$.
Once the five parameters $\beta$ and 
$\lambda$ are fixed, a one-to-one correspondence is established between 
$\alpha$ and the combination $(a_0^0-a_0^2)$ of the scattering lengths.
Instead of considering $\alpha$ as the main variable of the problem, it is
preferable to consider $(a_0^0-a_0^2)$ as the principal variable of interest,
since it has a direct physical meaning. We have thus considered a variation
of $(a_0^0-a_0^2)$ within the interval 0.250--0.370 which essentially covers
the domain of generalized $\chi PT$, including standard $\chi PT$. For each
given value of $(a_0^0-a_0^2)$, the other combination $(2a_0^0+a_0^2)$ is 
calculable through the low energy threshold parameter formulas \cite{kmsf}.
\par
The correction (\rf{e25}) to the decay width depends also on the
electromagnetic low energy constants $k^r$. Some of these have been
calculated in the $SU(3)\times SU(3)$ case in standard $\chi PT$ in Ref.
\cite{bu} and could be converted to the $SU(2)\times SU(2)$ case.
We have assumed that the combination of the $k^r$s present in the standard
$\chi PT$ case (the terms in Eq. (\rf{e21}) not proportional to
$(\alpha-\beta)$ or to $(\alpha-\beta)^2)$ keeps its numerical value also
in generalized $\chi PT$. As to the additional low energy constants (strong
and electromagnetic) appearing in generalized $\chi PT$, we have considered
them as uncertainties of the order of $1/(16\pi^2)$ and added them
quadratically. Furthermore, numerically $(\alpha)_0$ in Eq. (\rf{e25}) has
been taken equal to $\alpha$. Finally, the following numerical values have
been used: $F_{\pi}=93.2$ MeV, $m_{\pi^+}=139.57$ MeV, $m_{\pi^0}=134.97$
MeV. We have taken $m_{\pi}$, which appears in higher order corrective
terms, equal to $m_{\pi^+}$ and $\mu=m_{\rho}=770$ MeV.
[An error exists in the conversion formulas of Ref. \cite{jss} of the
combination of $k^r$s of the standard $\chi PT$ case, due to missing
factors coming from the strong interaction low energy constants $L^r$ and
from the difference of chiral limits of $F_{\pi}$ in the $SU(3)\times SU(3)$
and $SU(2)\times SU(2)$ cases. The correct formula has been given in Ref. 
\cite{gglr}. Numerical predictions already made are however almost
insensitive to this modification.]
\par
The uncertainty in the relative correction (\rf{e25}) was estimated in
Ref. \cite{jss}, in the standard $\chi PT$ case, to be of the order of 2\%
with respect to $\Gamma_0$. This uncertainty has been maintained here and the
additional one coming from the generalized $\chi PT$ coefficients added to
it. We have found that the uncertainty varies from 2\% (standard $\chi PT$
case) to 2.5\% (extreme case of generalized $\chi PT$).
\par
\begin{table}[htbp]
\begin{center}
\begin{tabular}{|ccccccc|}
\hline
$a_0^0-a_0^2$ & $\alpha$ & $a_0^0$ & $10a_0^2$ & 
$\tau_0\ (fs)$ & $(\Delta \tau/\tau_0)$ & $\tau\ (fs)$\\
\hline
0.250&0.77&0.205&$-0.448$&3.40&$-0.066$&3.20\\
0.254&0.89&0.211&$-0.431$&3.30&$-0.063$&3.11\\
0.258&1.02&0.217&$-0.413$&3.19&$-0.061$&3.03\\
0.262&1.13&0.222&$-0.397$&3.10&$-0.058$&2.94\\
0.270&1.37&0.234&$-0.363$&2.92&$-0.054$&2.78\\
0.290&1.95&0.262&$-0.281$&2.53&$-0.043$&2.44\\
0.300&2.23&0.276&$-0.241$&2.36&$-0.038$&2.29\\
0.310&2.51&0.290&$-0.201$&2.21&$-0.034$&2.16\\
0.320&2.78&0.304&$-0.161$&2.08&$-0.029$&2.03\\
0.330&3.05&0.318&$-0.122$&1.95&$-0.025$&1.92\\
0.340&3.31&0.332&$-0.083$&1.84&$-0.021$&1.81\\
0.350&3.57&0.346&$-0.044$&1.74&$-0.018$&1.72\\
0.360&3.83&0.360&$-0.006$&1.64&$-0.014$&1.63\\
0.370&4.09&0.373&$+0.032$&1.55&$-0.011$&1.55\\
\hline
\end{tabular}
\end{center}
\caption{The pionium lifetime $\tau$ as a function of $a_0^0-a_0^2$, with
the low energy constants $\beta$ and $\lambda$ fixed from the set I
solution of standard $\chi PT$ \cite{bcegs}. $\tau_0$ is the lifetime
obtained from the nonrelativistic formula.}
\lb{t1}
\end{table}
We present in Table \rf{t1} values of the pionium lifetime for
some typical values of $(a_0^0-a_0^2)$, as well as the corresponding
values of $\alpha$, $a_0^0$, $a_0^2$ and the relative correction 
corresponding to Eq. (\rf{e25}).
\par
We observe that the relative correction $|\Delta \tau/\tau_0|$ decreases
with increasing $\alpha$. This is mainly due to the fact that the tree level
correction tends to zero as $\alpha$ approaches the value 4. 
\par
In Fig. \rf{f1} we have represented the curve of the lifetime $\tau$ as a
function of the combination ($a_0^0-a_0^2$) of the scattering lengths 
(full line). The estimated uncertainties (2-2.5\%) are represented by
the band delineated by the dotted lines.
\par
\bfg
\vspace*{0.5 cm}
\begin{center}
\setlength{\unitlength}{0.240900pt}
\ifx\plotpoint\undefined\newsavebox{\plotpoint}\fi
\sbox{\plotpoint}{\rule[-0.200pt]{0.400pt}{0.400pt}}%
\begin{picture}(1800,1350)(0,0)
\font\gnuplot=cmr10 at 10pt
\gnuplot
\sbox{\plotpoint}{\rule[-0.200pt]{0.400pt}{0.400pt}}%
\put(181.0,163.0){\rule[-0.200pt]{4.818pt}{0.400pt}}
\put(161,163){\makebox(0,0)[r]{1}}
\put(1760.0,163.0){\rule[-0.200pt]{4.818pt}{0.400pt}}
\put(181.0,354.0){\rule[-0.200pt]{4.818pt}{0.400pt}}
\put(161,354){\makebox(0,0)[r]{1.5}}
\put(1760.0,354.0){\rule[-0.200pt]{4.818pt}{0.400pt}}
\put(181.0,545.0){\rule[-0.200pt]{4.818pt}{0.400pt}}
\put(161,545){\makebox(0,0)[r]{2}}
\put(1760.0,545.0){\rule[-0.200pt]{4.818pt}{0.400pt}}
\put(181.0,736.0){\rule[-0.200pt]{4.818pt}{0.400pt}}
\put(161,736){\makebox(0,0)[r]{2.5}}
\put(1760.0,736.0){\rule[-0.200pt]{4.818pt}{0.400pt}}
\put(181.0,927.0){\rule[-0.200pt]{4.818pt}{0.400pt}}
\put(161,927){\makebox(0,0)[r]{3}}
\put(1760.0,927.0){\rule[-0.200pt]{4.818pt}{0.400pt}}
\put(181.0,1118.0){\rule[-0.200pt]{4.818pt}{0.400pt}}
\put(161,1118){\makebox(0,0)[r]{3.5}}
\put(1760.0,1118.0){\rule[-0.200pt]{4.818pt}{0.400pt}}
\put(181.0,1309.0){\rule[-0.200pt]{4.818pt}{0.400pt}}
\put(161,1309){\makebox(0,0)[r]{4}}
\put(1760.0,1309.0){\rule[-0.200pt]{4.818pt}{0.400pt}}
\put(181.0,163.0){\rule[-0.200pt]{0.400pt}{4.818pt}}
\put(181,122){\makebox(0,0){0.25}}
\put(181.0,1289.0){\rule[-0.200pt]{0.400pt}{4.818pt}}
\put(314.0,163.0){\rule[-0.200pt]{0.400pt}{4.818pt}}
\put(314,122){\makebox(0,0){0.26}}
\put(314.0,1289.0){\rule[-0.200pt]{0.400pt}{4.818pt}}
\put(448.0,163.0){\rule[-0.200pt]{0.400pt}{4.818pt}}
\put(448,122){\makebox(0,0){0.27}}
\put(448.0,1289.0){\rule[-0.200pt]{0.400pt}{4.818pt}}
\put(581.0,163.0){\rule[-0.200pt]{0.400pt}{4.818pt}}
\put(581,122){\makebox(0,0){0.28}}
\put(581.0,1289.0){\rule[-0.200pt]{0.400pt}{4.818pt}}
\put(714.0,163.0){\rule[-0.200pt]{0.400pt}{4.818pt}}
\put(714,122){\makebox(0,0){0.29}}
\put(714.0,1289.0){\rule[-0.200pt]{0.400pt}{4.818pt}}
\put(847.0,163.0){\rule[-0.200pt]{0.400pt}{4.818pt}}
\put(847,122){\makebox(0,0){0.3}}
\put(847.0,1289.0){\rule[-0.200pt]{0.400pt}{4.818pt}}
\put(981.0,163.0){\rule[-0.200pt]{0.400pt}{4.818pt}}
\put(981,122){\makebox(0,0){0.31}}
\put(981.0,1289.0){\rule[-0.200pt]{0.400pt}{4.818pt}}
\put(1114.0,163.0){\rule[-0.200pt]{0.400pt}{4.818pt}}
\put(1114,122){\makebox(0,0){0.32}}
\put(1114.0,1289.0){\rule[-0.200pt]{0.400pt}{4.818pt}}
\put(1247.0,163.0){\rule[-0.200pt]{0.400pt}{4.818pt}}
\put(1247,122){\makebox(0,0){0.33}}
\put(1247.0,1289.0){\rule[-0.200pt]{0.400pt}{4.818pt}}
\put(1380.0,163.0){\rule[-0.200pt]{0.400pt}{4.818pt}}
\put(1380,122){\makebox(0,0){0.34}}
\put(1380.0,1289.0){\rule[-0.200pt]{0.400pt}{4.818pt}}
\put(1514.0,163.0){\rule[-0.200pt]{0.400pt}{4.818pt}}
\put(1514,122){\makebox(0,0){0.35}}
\put(1514.0,1289.0){\rule[-0.200pt]{0.400pt}{4.818pt}}
\put(1647.0,163.0){\rule[-0.200pt]{0.400pt}{4.818pt}}
\put(1647,122){\makebox(0,0){0.36}}
\put(1647.0,1289.0){\rule[-0.200pt]{0.400pt}{4.818pt}}
\put(1780.0,163.0){\rule[-0.200pt]{0.400pt}{4.818pt}}
\put(1780,122){\makebox(0,0){0.37}}
\put(1780.0,1289.0){\rule[-0.200pt]{0.400pt}{4.818pt}}
\put(181.0,163.0){\rule[-0.200pt]{385.199pt}{0.400pt}}
\put(1780.0,163.0){\rule[-0.200pt]{0.400pt}{276.071pt}}
\put(181.0,1309.0){\rule[-0.200pt]{385.199pt}{0.400pt}}
\put(82,1216){\makebox(0,0){$\tau\ (fs)$}}
\put(980,61){\makebox(0,0){$a_0^0-a_0^2$}}
\put(181.0,163.0){\rule[-0.200pt]{0.400pt}{276.071pt}}
\multiput(181.00,1003.92)(0.808,-0.491){17}{\rule{0.740pt}{0.118pt}}
\multiput(181.00,1004.17)(14.464,-10.000){2}{\rule{0.370pt}{0.400pt}}
\multiput(197.00,993.92)(0.732,-0.492){19}{\rule{0.682pt}{0.118pt}}
\multiput(197.00,994.17)(14.585,-11.000){2}{\rule{0.341pt}{0.400pt}}
\multiput(213.00,982.92)(0.732,-0.492){19}{\rule{0.682pt}{0.118pt}}
\multiput(213.00,983.17)(14.585,-11.000){2}{\rule{0.341pt}{0.400pt}}
\multiput(229.00,971.92)(0.860,-0.491){17}{\rule{0.780pt}{0.118pt}}
\multiput(229.00,972.17)(15.381,-10.000){2}{\rule{0.390pt}{0.400pt}}
\multiput(246.00,961.92)(0.808,-0.491){17}{\rule{0.740pt}{0.118pt}}
\multiput(246.00,962.17)(14.464,-10.000){2}{\rule{0.370pt}{0.400pt}}
\multiput(262.00,951.92)(0.808,-0.491){17}{\rule{0.740pt}{0.118pt}}
\multiput(262.00,952.17)(14.464,-10.000){2}{\rule{0.370pt}{0.400pt}}
\multiput(278.00,941.92)(0.808,-0.491){17}{\rule{0.740pt}{0.118pt}}
\multiput(278.00,942.17)(14.464,-10.000){2}{\rule{0.370pt}{0.400pt}}
\multiput(294.00,931.92)(0.808,-0.491){17}{\rule{0.740pt}{0.118pt}}
\multiput(294.00,932.17)(14.464,-10.000){2}{\rule{0.370pt}{0.400pt}}
\multiput(310.00,921.92)(0.808,-0.491){17}{\rule{0.740pt}{0.118pt}}
\multiput(310.00,922.17)(14.464,-10.000){2}{\rule{0.370pt}{0.400pt}}
\multiput(326.00,911.93)(0.961,-0.489){15}{\rule{0.856pt}{0.118pt}}
\multiput(326.00,912.17)(15.224,-9.000){2}{\rule{0.428pt}{0.400pt}}
\multiput(343.00,902.92)(0.808,-0.491){17}{\rule{0.740pt}{0.118pt}}
\multiput(343.00,903.17)(14.464,-10.000){2}{\rule{0.370pt}{0.400pt}}
\multiput(359.00,892.93)(0.902,-0.489){15}{\rule{0.811pt}{0.118pt}}
\multiput(359.00,893.17)(14.316,-9.000){2}{\rule{0.406pt}{0.400pt}}
\multiput(375.00,883.93)(0.902,-0.489){15}{\rule{0.811pt}{0.118pt}}
\multiput(375.00,884.17)(14.316,-9.000){2}{\rule{0.406pt}{0.400pt}}
\multiput(391.00,874.92)(0.808,-0.491){17}{\rule{0.740pt}{0.118pt}}
\multiput(391.00,875.17)(14.464,-10.000){2}{\rule{0.370pt}{0.400pt}}
\multiput(407.00,864.93)(0.902,-0.489){15}{\rule{0.811pt}{0.118pt}}
\multiput(407.00,865.17)(14.316,-9.000){2}{\rule{0.406pt}{0.400pt}}
\multiput(423.00,855.93)(1.022,-0.488){13}{\rule{0.900pt}{0.117pt}}
\multiput(423.00,856.17)(14.132,-8.000){2}{\rule{0.450pt}{0.400pt}}
\multiput(439.00,847.93)(0.961,-0.489){15}{\rule{0.856pt}{0.118pt}}
\multiput(439.00,848.17)(15.224,-9.000){2}{\rule{0.428pt}{0.400pt}}
\multiput(456.00,838.93)(0.902,-0.489){15}{\rule{0.811pt}{0.118pt}}
\multiput(456.00,839.17)(14.316,-9.000){2}{\rule{0.406pt}{0.400pt}}
\multiput(472.00,829.93)(1.022,-0.488){13}{\rule{0.900pt}{0.117pt}}
\multiput(472.00,830.17)(14.132,-8.000){2}{\rule{0.450pt}{0.400pt}}
\multiput(488.00,821.93)(0.902,-0.489){15}{\rule{0.811pt}{0.118pt}}
\multiput(488.00,822.17)(14.316,-9.000){2}{\rule{0.406pt}{0.400pt}}
\multiput(504.00,812.93)(1.022,-0.488){13}{\rule{0.900pt}{0.117pt}}
\multiput(504.00,813.17)(14.132,-8.000){2}{\rule{0.450pt}{0.400pt}}
\multiput(520.00,804.93)(0.902,-0.489){15}{\rule{0.811pt}{0.118pt}}
\multiput(520.00,805.17)(14.316,-9.000){2}{\rule{0.406pt}{0.400pt}}
\multiput(536.00,795.93)(1.088,-0.488){13}{\rule{0.950pt}{0.117pt}}
\multiput(536.00,796.17)(15.028,-8.000){2}{\rule{0.475pt}{0.400pt}}
\multiput(553.00,787.93)(1.022,-0.488){13}{\rule{0.900pt}{0.117pt}}
\multiput(553.00,788.17)(14.132,-8.000){2}{\rule{0.450pt}{0.400pt}}
\multiput(569.00,779.93)(1.022,-0.488){13}{\rule{0.900pt}{0.117pt}}
\multiput(569.00,780.17)(14.132,-8.000){2}{\rule{0.450pt}{0.400pt}}
\multiput(585.00,771.93)(1.022,-0.488){13}{\rule{0.900pt}{0.117pt}}
\multiput(585.00,772.17)(14.132,-8.000){2}{\rule{0.450pt}{0.400pt}}
\multiput(601.00,763.93)(1.179,-0.485){11}{\rule{1.014pt}{0.117pt}}
\multiput(601.00,764.17)(13.895,-7.000){2}{\rule{0.507pt}{0.400pt}}
\multiput(617.00,756.93)(1.022,-0.488){13}{\rule{0.900pt}{0.117pt}}
\multiput(617.00,757.17)(14.132,-8.000){2}{\rule{0.450pt}{0.400pt}}
\multiput(633.00,748.93)(1.022,-0.488){13}{\rule{0.900pt}{0.117pt}}
\multiput(633.00,749.17)(14.132,-8.000){2}{\rule{0.450pt}{0.400pt}}
\multiput(649.00,740.93)(1.255,-0.485){11}{\rule{1.071pt}{0.117pt}}
\multiput(649.00,741.17)(14.776,-7.000){2}{\rule{0.536pt}{0.400pt}}
\multiput(666.00,733.93)(1.022,-0.488){13}{\rule{0.900pt}{0.117pt}}
\multiput(666.00,734.17)(14.132,-8.000){2}{\rule{0.450pt}{0.400pt}}
\multiput(682.00,725.93)(1.179,-0.485){11}{\rule{1.014pt}{0.117pt}}
\multiput(682.00,726.17)(13.895,-7.000){2}{\rule{0.507pt}{0.400pt}}
\multiput(698.00,718.93)(1.179,-0.485){11}{\rule{1.014pt}{0.117pt}}
\multiput(698.00,719.17)(13.895,-7.000){2}{\rule{0.507pt}{0.400pt}}
\multiput(714.00,711.93)(1.179,-0.485){11}{\rule{1.014pt}{0.117pt}}
\multiput(714.00,712.17)(13.895,-7.000){2}{\rule{0.507pt}{0.400pt}}
\multiput(730.00,704.93)(1.179,-0.485){11}{\rule{1.014pt}{0.117pt}}
\multiput(730.00,705.17)(13.895,-7.000){2}{\rule{0.507pt}{0.400pt}}
\multiput(746.00,697.93)(1.255,-0.485){11}{\rule{1.071pt}{0.117pt}}
\multiput(746.00,698.17)(14.776,-7.000){2}{\rule{0.536pt}{0.400pt}}
\multiput(763.00,690.93)(1.179,-0.485){11}{\rule{1.014pt}{0.117pt}}
\multiput(763.00,691.17)(13.895,-7.000){2}{\rule{0.507pt}{0.400pt}}
\multiput(779.00,683.93)(1.179,-0.485){11}{\rule{1.014pt}{0.117pt}}
\multiput(779.00,684.17)(13.895,-7.000){2}{\rule{0.507pt}{0.400pt}}
\multiput(795.00,676.93)(1.179,-0.485){11}{\rule{1.014pt}{0.117pt}}
\multiput(795.00,677.17)(13.895,-7.000){2}{\rule{0.507pt}{0.400pt}}
\multiput(811.00,669.93)(1.179,-0.485){11}{\rule{1.014pt}{0.117pt}}
\multiput(811.00,670.17)(13.895,-7.000){2}{\rule{0.507pt}{0.400pt}}
\multiput(827.00,662.93)(1.395,-0.482){9}{\rule{1.167pt}{0.116pt}}
\multiput(827.00,663.17)(13.579,-6.000){2}{\rule{0.583pt}{0.400pt}}
\multiput(843.00,656.93)(1.255,-0.485){11}{\rule{1.071pt}{0.117pt}}
\multiput(843.00,657.17)(14.776,-7.000){2}{\rule{0.536pt}{0.400pt}}
\multiput(860.00,649.93)(1.395,-0.482){9}{\rule{1.167pt}{0.116pt}}
\multiput(860.00,650.17)(13.579,-6.000){2}{\rule{0.583pt}{0.400pt}}
\multiput(876.00,643.93)(1.179,-0.485){11}{\rule{1.014pt}{0.117pt}}
\multiput(876.00,644.17)(13.895,-7.000){2}{\rule{0.507pt}{0.400pt}}
\multiput(892.00,636.93)(1.395,-0.482){9}{\rule{1.167pt}{0.116pt}}
\multiput(892.00,637.17)(13.579,-6.000){2}{\rule{0.583pt}{0.400pt}}
\multiput(908.00,630.93)(1.395,-0.482){9}{\rule{1.167pt}{0.116pt}}
\multiput(908.00,631.17)(13.579,-6.000){2}{\rule{0.583pt}{0.400pt}}
\multiput(924.00,624.93)(1.395,-0.482){9}{\rule{1.167pt}{0.116pt}}
\multiput(924.00,625.17)(13.579,-6.000){2}{\rule{0.583pt}{0.400pt}}
\multiput(940.00,618.93)(1.179,-0.485){11}{\rule{1.014pt}{0.117pt}}
\multiput(940.00,619.17)(13.895,-7.000){2}{\rule{0.507pt}{0.400pt}}
\multiput(956.00,611.93)(1.485,-0.482){9}{\rule{1.233pt}{0.116pt}}
\multiput(956.00,612.17)(14.440,-6.000){2}{\rule{0.617pt}{0.400pt}}
\multiput(973.00,605.93)(1.395,-0.482){9}{\rule{1.167pt}{0.116pt}}
\multiput(973.00,606.17)(13.579,-6.000){2}{\rule{0.583pt}{0.400pt}}
\multiput(989.00,599.93)(1.395,-0.482){9}{\rule{1.167pt}{0.116pt}}
\multiput(989.00,600.17)(13.579,-6.000){2}{\rule{0.583pt}{0.400pt}}
\multiput(1005.00,593.93)(1.712,-0.477){7}{\rule{1.380pt}{0.115pt}}
\multiput(1005.00,594.17)(13.136,-5.000){2}{\rule{0.690pt}{0.400pt}}
\multiput(1021.00,588.93)(1.395,-0.482){9}{\rule{1.167pt}{0.116pt}}
\multiput(1021.00,589.17)(13.579,-6.000){2}{\rule{0.583pt}{0.400pt}}
\multiput(1037.00,582.93)(1.395,-0.482){9}{\rule{1.167pt}{0.116pt}}
\multiput(1037.00,583.17)(13.579,-6.000){2}{\rule{0.583pt}{0.400pt}}
\multiput(1053.00,576.93)(1.485,-0.482){9}{\rule{1.233pt}{0.116pt}}
\multiput(1053.00,577.17)(14.440,-6.000){2}{\rule{0.617pt}{0.400pt}}
\multiput(1070.00,570.93)(1.712,-0.477){7}{\rule{1.380pt}{0.115pt}}
\multiput(1070.00,571.17)(13.136,-5.000){2}{\rule{0.690pt}{0.400pt}}
\multiput(1086.00,565.93)(1.395,-0.482){9}{\rule{1.167pt}{0.116pt}}
\multiput(1086.00,566.17)(13.579,-6.000){2}{\rule{0.583pt}{0.400pt}}
\multiput(1102.00,559.93)(1.712,-0.477){7}{\rule{1.380pt}{0.115pt}}
\multiput(1102.00,560.17)(13.136,-5.000){2}{\rule{0.690pt}{0.400pt}}
\multiput(1118.00,554.93)(1.395,-0.482){9}{\rule{1.167pt}{0.116pt}}
\multiput(1118.00,555.17)(13.579,-6.000){2}{\rule{0.583pt}{0.400pt}}
\multiput(1134.00,548.93)(1.712,-0.477){7}{\rule{1.380pt}{0.115pt}}
\multiput(1134.00,549.17)(13.136,-5.000){2}{\rule{0.690pt}{0.400pt}}
\multiput(1150.00,543.93)(1.712,-0.477){7}{\rule{1.380pt}{0.115pt}}
\multiput(1150.00,544.17)(13.136,-5.000){2}{\rule{0.690pt}{0.400pt}}
\multiput(1166.00,538.93)(1.485,-0.482){9}{\rule{1.233pt}{0.116pt}}
\multiput(1166.00,539.17)(14.440,-6.000){2}{\rule{0.617pt}{0.400pt}}
\multiput(1183.00,532.93)(1.712,-0.477){7}{\rule{1.380pt}{0.115pt}}
\multiput(1183.00,533.17)(13.136,-5.000){2}{\rule{0.690pt}{0.400pt}}
\multiput(1199.00,527.93)(1.712,-0.477){7}{\rule{1.380pt}{0.115pt}}
\multiput(1199.00,528.17)(13.136,-5.000){2}{\rule{0.690pt}{0.400pt}}
\multiput(1215.00,522.93)(1.712,-0.477){7}{\rule{1.380pt}{0.115pt}}
\multiput(1215.00,523.17)(13.136,-5.000){2}{\rule{0.690pt}{0.400pt}}
\multiput(1231.00,517.93)(1.712,-0.477){7}{\rule{1.380pt}{0.115pt}}
\multiput(1231.00,518.17)(13.136,-5.000){2}{\rule{0.690pt}{0.400pt}}
\multiput(1247.00,512.93)(1.712,-0.477){7}{\rule{1.380pt}{0.115pt}}
\multiput(1247.00,513.17)(13.136,-5.000){2}{\rule{0.690pt}{0.400pt}}
\multiput(1263.00,507.93)(1.823,-0.477){7}{\rule{1.460pt}{0.115pt}}
\multiput(1263.00,508.17)(13.970,-5.000){2}{\rule{0.730pt}{0.400pt}}
\multiput(1280.00,502.93)(1.712,-0.477){7}{\rule{1.380pt}{0.115pt}}
\multiput(1280.00,503.17)(13.136,-5.000){2}{\rule{0.690pt}{0.400pt}}
\multiput(1296.00,497.93)(1.712,-0.477){7}{\rule{1.380pt}{0.115pt}}
\multiput(1296.00,498.17)(13.136,-5.000){2}{\rule{0.690pt}{0.400pt}}
\multiput(1312.00,492.93)(1.712,-0.477){7}{\rule{1.380pt}{0.115pt}}
\multiput(1312.00,493.17)(13.136,-5.000){2}{\rule{0.690pt}{0.400pt}}
\multiput(1328.00,487.94)(2.236,-0.468){5}{\rule{1.700pt}{0.113pt}}
\multiput(1328.00,488.17)(12.472,-4.000){2}{\rule{0.850pt}{0.400pt}}
\multiput(1344.00,483.93)(1.712,-0.477){7}{\rule{1.380pt}{0.115pt}}
\multiput(1344.00,484.17)(13.136,-5.000){2}{\rule{0.690pt}{0.400pt}}
\multiput(1360.00,478.93)(1.823,-0.477){7}{\rule{1.460pt}{0.115pt}}
\multiput(1360.00,479.17)(13.970,-5.000){2}{\rule{0.730pt}{0.400pt}}
\multiput(1377.00,473.94)(2.236,-0.468){5}{\rule{1.700pt}{0.113pt}}
\multiput(1377.00,474.17)(12.472,-4.000){2}{\rule{0.850pt}{0.400pt}}
\multiput(1393.00,469.93)(1.712,-0.477){7}{\rule{1.380pt}{0.115pt}}
\multiput(1393.00,470.17)(13.136,-5.000){2}{\rule{0.690pt}{0.400pt}}
\multiput(1409.00,464.93)(1.712,-0.477){7}{\rule{1.380pt}{0.115pt}}
\multiput(1409.00,465.17)(13.136,-5.000){2}{\rule{0.690pt}{0.400pt}}
\multiput(1425.00,459.94)(2.236,-0.468){5}{\rule{1.700pt}{0.113pt}}
\multiput(1425.00,460.17)(12.472,-4.000){2}{\rule{0.850pt}{0.400pt}}
\multiput(1441.00,455.94)(2.236,-0.468){5}{\rule{1.700pt}{0.113pt}}
\multiput(1441.00,456.17)(12.472,-4.000){2}{\rule{0.850pt}{0.400pt}}
\multiput(1457.00,451.93)(1.712,-0.477){7}{\rule{1.380pt}{0.115pt}}
\multiput(1457.00,452.17)(13.136,-5.000){2}{\rule{0.690pt}{0.400pt}}
\multiput(1473.00,446.94)(2.382,-0.468){5}{\rule{1.800pt}{0.113pt}}
\multiput(1473.00,447.17)(13.264,-4.000){2}{\rule{0.900pt}{0.400pt}}
\multiput(1490.00,442.94)(2.236,-0.468){5}{\rule{1.700pt}{0.113pt}}
\multiput(1490.00,443.17)(12.472,-4.000){2}{\rule{0.850pt}{0.400pt}}
\multiput(1506.00,438.93)(1.712,-0.477){7}{\rule{1.380pt}{0.115pt}}
\multiput(1506.00,439.17)(13.136,-5.000){2}{\rule{0.690pt}{0.400pt}}
\multiput(1522.00,433.94)(2.236,-0.468){5}{\rule{1.700pt}{0.113pt}}
\multiput(1522.00,434.17)(12.472,-4.000){2}{\rule{0.850pt}{0.400pt}}
\multiput(1538.00,429.94)(2.236,-0.468){5}{\rule{1.700pt}{0.113pt}}
\multiput(1538.00,430.17)(12.472,-4.000){2}{\rule{0.850pt}{0.400pt}}
\multiput(1554.00,425.94)(2.236,-0.468){5}{\rule{1.700pt}{0.113pt}}
\multiput(1554.00,426.17)(12.472,-4.000){2}{\rule{0.850pt}{0.400pt}}
\multiput(1570.00,421.94)(2.382,-0.468){5}{\rule{1.800pt}{0.113pt}}
\multiput(1570.00,422.17)(13.264,-4.000){2}{\rule{0.900pt}{0.400pt}}
\multiput(1587.00,417.94)(2.236,-0.468){5}{\rule{1.700pt}{0.113pt}}
\multiput(1587.00,418.17)(12.472,-4.000){2}{\rule{0.850pt}{0.400pt}}
\multiput(1603.00,413.94)(2.236,-0.468){5}{\rule{1.700pt}{0.113pt}}
\multiput(1603.00,414.17)(12.472,-4.000){2}{\rule{0.850pt}{0.400pt}}
\multiput(1619.00,409.94)(2.236,-0.468){5}{\rule{1.700pt}{0.113pt}}
\multiput(1619.00,410.17)(12.472,-4.000){2}{\rule{0.850pt}{0.400pt}}
\multiput(1635.00,405.94)(2.236,-0.468){5}{\rule{1.700pt}{0.113pt}}
\multiput(1635.00,406.17)(12.472,-4.000){2}{\rule{0.850pt}{0.400pt}}
\multiput(1651.00,401.94)(2.236,-0.468){5}{\rule{1.700pt}{0.113pt}}
\multiput(1651.00,402.17)(12.472,-4.000){2}{\rule{0.850pt}{0.400pt}}
\multiput(1667.00,397.94)(2.382,-0.468){5}{\rule{1.800pt}{0.113pt}}
\multiput(1667.00,398.17)(13.264,-4.000){2}{\rule{0.900pt}{0.400pt}}
\multiput(1684.00,393.94)(2.236,-0.468){5}{\rule{1.700pt}{0.113pt}}
\multiput(1684.00,394.17)(12.472,-4.000){2}{\rule{0.850pt}{0.400pt}}
\multiput(1700.00,389.94)(2.236,-0.468){5}{\rule{1.700pt}{0.113pt}}
\multiput(1700.00,390.17)(12.472,-4.000){2}{\rule{0.850pt}{0.400pt}}
\multiput(1716.00,385.94)(2.236,-0.468){5}{\rule{1.700pt}{0.113pt}}
\multiput(1716.00,386.17)(12.472,-4.000){2}{\rule{0.850pt}{0.400pt}}
\multiput(1732.00,381.95)(3.365,-0.447){3}{\rule{2.233pt}{0.108pt}}
\multiput(1732.00,382.17)(11.365,-3.000){2}{\rule{1.117pt}{0.400pt}}
\multiput(1748.00,378.94)(2.236,-0.468){5}{\rule{1.700pt}{0.113pt}}
\multiput(1748.00,379.17)(12.472,-4.000){2}{\rule{0.850pt}{0.400pt}}
\multiput(1764.00,374.94)(2.236,-0.468){5}{\rule{1.700pt}{0.113pt}}
\multiput(1764.00,375.17)(12.472,-4.000){2}{\rule{0.850pt}{0.400pt}}
\sbox{\plotpoint}{\rule[-0.500pt]{1.000pt}{1.000pt}}%
\put(181.00,1031.00){\usebox{\plotpoint}}
\put(198.10,1019.24){\usebox{\plotpoint}}
\put(215.21,1007.48){\usebox{\plotpoint}}
\put(232.46,995.96){\usebox{\plotpoint}}
\put(250.16,985.14){\usebox{\plotpoint}}
\put(267.42,973.61){\usebox{\plotpoint}}
\put(285.02,962.61){\usebox{\plotpoint}}
\put(302.62,951.61){\usebox{\plotpoint}}
\put(320.22,940.61){\usebox{\plotpoint}}
\put(338.01,929.93){\usebox{\plotpoint}}
\put(356.05,919.66){\usebox{\plotpoint}}
\put(373.73,908.79){\usebox{\plotpoint}}
\put(391.76,898.52){\usebox{\plotpoint}}
\put(409.43,887.63){\usebox{\plotpoint}}
\put(427.52,877.46){\usebox{\plotpoint}}
\put(445.70,867.45){\usebox{\plotpoint}}
\put(464.14,857.93){\usebox{\plotpoint}}
\put(482.43,848.13){\usebox{\plotpoint}}
\put(500.52,837.96){\usebox{\plotpoint}}
\put(519.00,828.50){\usebox{\plotpoint}}
\put(537.15,818.46){\usebox{\plotpoint}}
\put(555.90,809.55){\usebox{\plotpoint}}
\put(574.46,800.27){\usebox{\plotpoint}}
\put(593.03,790.99){\usebox{\plotpoint}}
\put(611.59,781.70){\usebox{\plotpoint}}
\put(630.16,772.42){\usebox{\plotpoint}}
\put(648.72,763.14){\usebox{\plotpoint}}
\put(667.84,755.08){\usebox{\plotpoint}}
\put(686.51,746.03){\usebox{\plotpoint}}
\put(705.35,737.33){\usebox{\plotpoint}}
\put(724.15,728.56){\usebox{\plotpoint}}
\put(743.17,720.24){\usebox{\plotpoint}}
\put(762.34,712.27){\usebox{\plotpoint}}
\put(781.01,703.25){\usebox{\plotpoint}}
\put(800.33,695.67){\usebox{\plotpoint}}
\put(819.34,687.35){\usebox{\plotpoint}}
\put(838.36,679.03){\usebox{\plotpoint}}
\put(857.51,671.03){\usebox{\plotpoint}}
\put(876.89,663.61){\usebox{\plotpoint}}
\put(895.99,655.50){\usebox{\plotpoint}}
\put(915.43,648.22){\usebox{\plotpoint}}
\put(934.63,640.35){\usebox{\plotpoint}}
\put(953.94,632.77){\usebox{\plotpoint}}
\put(973.50,625.81){\usebox{\plotpoint}}
\put(992.93,618.53){\usebox{\plotpoint}}
\put(1012.36,611.24){\usebox{\plotpoint}}
\put(1031.80,603.95){\usebox{\plotpoint}}
\put(1051.23,596.66){\usebox{\plotpoint}}
\put(1070.79,589.71){\usebox{\plotpoint}}
\put(1090.30,582.66){\usebox{\plotpoint}}
\put(1109.96,576.02){\usebox{\plotpoint}}
\put(1129.61,569.37){\usebox{\plotpoint}}
\put(1149.13,562.33){\usebox{\plotpoint}}
\put(1168.89,555.98){\usebox{\plotpoint}}
\put(1188.53,549.27){\usebox{\plotpoint}}
\put(1208.34,543.08){\usebox{\plotpoint}}
\put(1228.15,536.89){\usebox{\plotpoint}}
\put(1247.94,530.65){\usebox{\plotpoint}}
\put(1267.48,523.68){\usebox{\plotpoint}}
\put(1287.36,517.70){\usebox{\plotpoint}}
\put(1307.17,511.51){\usebox{\plotpoint}}
\put(1327.22,506.19){\usebox{\plotpoint}}
\put(1347.05,500.05){\usebox{\plotpoint}}
\put(1366.89,493.97){\usebox{\plotpoint}}
\put(1386.76,487.95){\usebox{\plotpoint}}
\put(1406.79,482.55){\usebox{\plotpoint}}
\put(1426.66,476.58){\usebox{\plotpoint}}
\put(1446.71,471.22){\usebox{\plotpoint}}
\put(1466.67,465.58){\usebox{\plotpoint}}
\put(1486.65,459.98){\usebox{\plotpoint}}
\put(1506.75,454.81){\usebox{\plotpoint}}
\put(1526.81,449.50){\usebox{\plotpoint}}
\put(1546.76,443.81){\usebox{\plotpoint}}
\put(1566.90,438.78){\usebox{\plotpoint}}
\put(1587.09,433.97){\usebox{\plotpoint}}
\put(1606.96,428.01){\usebox{\plotpoint}}
\put(1627.10,422.98){\usebox{\plotpoint}}
\put(1647.24,417.94){\usebox{\plotpoint}}
\put(1667.37,412.91){\usebox{\plotpoint}}
\put(1687.56,408.11){\usebox{\plotpoint}}
\put(1707.70,403.08){\usebox{\plotpoint}}
\put(1727.99,398.75){\usebox{\plotpoint}}
\put(1748.18,393.96){\usebox{\plotpoint}}
\put(1768.31,388.92){\usebox{\plotpoint}}
\put(1780,386){\usebox{\plotpoint}}
\put(181.00,980.00){\usebox{\plotpoint}}
\put(198.14,968.29){\usebox{\plotpoint}}
\put(215.66,957.17){\usebox{\plotpoint}}
\put(232.94,945.69){\usebox{\plotpoint}}
\put(250.75,935.03){\usebox{\plotpoint}}
\put(268.35,924.03){\usebox{\plotpoint}}
\put(286.17,913.40){\usebox{\plotpoint}}
\put(303.98,902.76){\usebox{\plotpoint}}
\put(321.58,891.76){\usebox{\plotpoint}}
\put(339.74,881.73){\usebox{\plotpoint}}
\put(357.87,871.63){\usebox{\plotpoint}}
\put(375.52,860.71){\usebox{\plotpoint}}
\put(393.61,850.53){\usebox{\plotpoint}}
\put(411.82,840.59){\usebox{\plotpoint}}
\put(430.20,830.95){\usebox{\plotpoint}}
\put(448.42,821.01){\usebox{\plotpoint}}
\put(466.89,811.55){\usebox{\plotpoint}}
\put(485.11,801.62){\usebox{\plotpoint}}
\put(503.60,792.20){\usebox{\plotpoint}}
\put(522.16,782.92){\usebox{\plotpoint}}
\put(540.78,773.75){\usebox{\plotpoint}}
\put(559.49,764.76){\usebox{\plotpoint}}
\put(578.05,755.47){\usebox{\plotpoint}}
\put(596.62,746.19){\usebox{\plotpoint}}
\put(615.18,736.91){\usebox{\plotpoint}}
\put(634.13,728.44){\usebox{\plotpoint}}
\put(652.81,719.43){\usebox{\plotpoint}}
\put(671.95,711.40){\usebox{\plotpoint}}
\put(690.97,703.08){\usebox{\plotpoint}}
\put(709.70,694.15){\usebox{\plotpoint}}
\put(728.61,685.61){\usebox{\plotpoint}}
\put(747.67,677.41){\usebox{\plotpoint}}
\put(767.12,670.20){\usebox{\plotpoint}}
\put(786.14,661.88){\usebox{\plotpoint}}
\put(805.15,653.56){\usebox{\plotpoint}}
\put(824.46,645.95){\usebox{\plotpoint}}
\put(843.54,637.81){\usebox{\plotpoint}}
\put(863.09,630.84){\usebox{\plotpoint}}
\put(882.39,623.21){\usebox{\plotpoint}}
\put(901.61,615.40){\usebox{\plotpoint}}
\put(921.04,608.11){\usebox{\plotpoint}}
\put(940.48,600.82){\usebox{\plotpoint}}
\put(959.94,593.61){\usebox{\plotpoint}}
\put(979.47,586.58){\usebox{\plotpoint}}
\put(998.90,579.29){\usebox{\plotpoint}}
\put(1018.59,572.75){\usebox{\plotpoint}}
\put(1038.07,565.60){\usebox{\plotpoint}}
\put(1057.62,558.64){\usebox{\plotpoint}}
\put(1077.35,552.24){\usebox{\plotpoint}}
\put(1096.99,545.57){\usebox{\plotpoint}}
\put(1116.52,538.55){\usebox{\plotpoint}}
\put(1136.30,532.28){\usebox{\plotpoint}}
\put(1156.00,525.75){\usebox{\plotpoint}}
\put(1175.66,519.16){\usebox{\plotpoint}}
\put(1195.51,513.09){\usebox{\plotpoint}}
\put(1215.32,506.90){\usebox{\plotpoint}}
\put(1235.13,500.71){\usebox{\plotpoint}}
\put(1254.94,494.52){\usebox{\plotpoint}}
\put(1274.81,488.53){\usebox{\plotpoint}}
\put(1294.89,483.28){\usebox{\plotpoint}}
\put(1314.72,477.15){\usebox{\plotpoint}}
\put(1334.53,470.96){\usebox{\plotpoint}}
\put(1354.51,465.37){\usebox{\plotpoint}}
\put(1374.48,459.74){\usebox{\plotpoint}}
\put(1394.57,454.51){\usebox{\plotpoint}}
\put(1414.46,448.63){\usebox{\plotpoint}}
\put(1434.45,443.05){\usebox{\plotpoint}}
\put(1454.47,437.63){\usebox{\plotpoint}}
\put(1474.59,432.53){\usebox{\plotpoint}}
\put(1494.55,426.86){\usebox{\plotpoint}}
\put(1514.69,421.83){\usebox{\plotpoint}}
\put(1534.83,416.79){\usebox{\plotpoint}}
\put(1554.96,411.76){\usebox{\plotpoint}}
\put(1575.11,406.80){\usebox{\plotpoint}}
\put(1595.29,401.93){\usebox{\plotpoint}}
\put(1615.43,396.89){\usebox{\plotpoint}}
\put(1635.56,391.86){\usebox{\plotpoint}}
\put(1655.70,386.83){\usebox{\plotpoint}}
\put(1675.86,381.91){\usebox{\plotpoint}}
\put(1696.18,377.72){\usebox{\plotpoint}}
\put(1716.37,372.91){\usebox{\plotpoint}}
\put(1736.56,368.14){\usebox{\plotpoint}}
\put(1756.85,363.79){\usebox{\plotpoint}}
\put(1776.98,358.75){\usebox{\plotpoint}}
\put(1780,358){\usebox{\plotpoint}}
\end{picture}
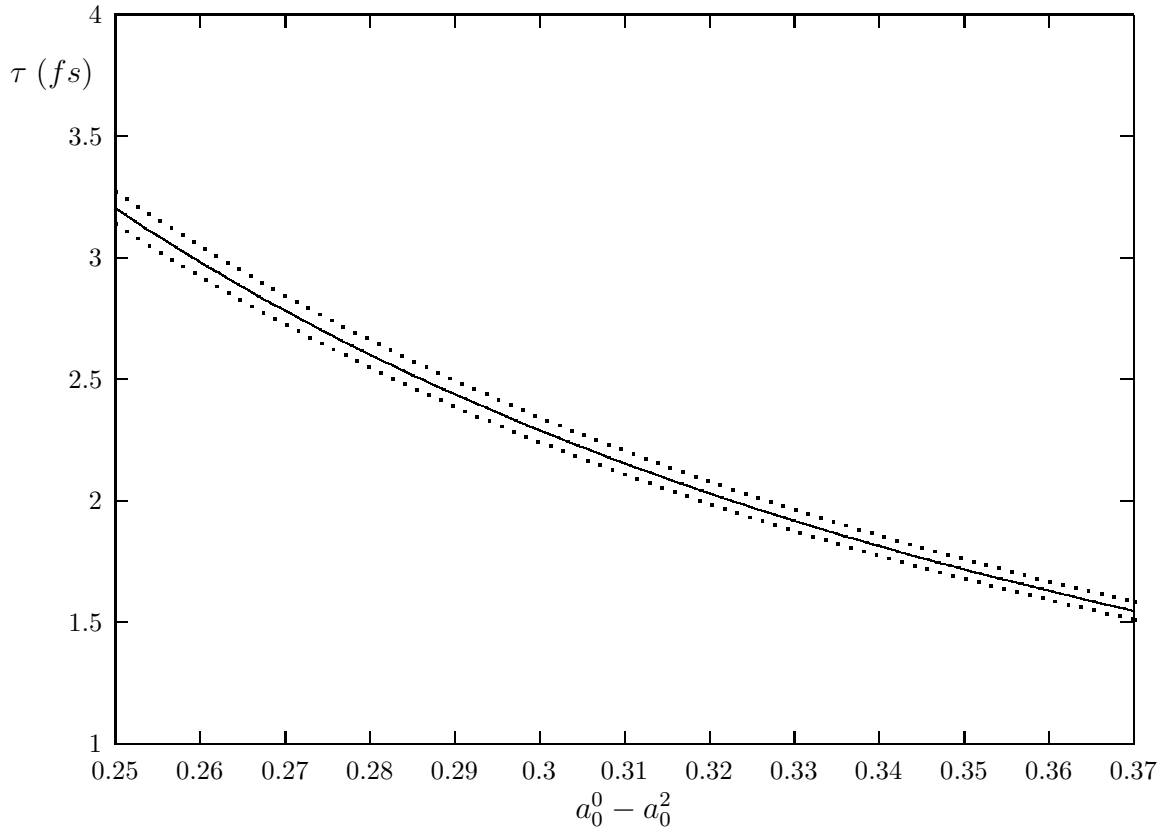
\caption{The pionium lifetime as a function of the combination
$(a_0^0-a_0^2)$ of the $S$-wave scattering lengths (full line). 
The band delineated by the dotted lines takes into account the estimated
uncertainties (2-2.5\%).}
\lb{f1}
\end{center}
\efg
The theoretical value of the lifetime depends also on the uncertainties
on the strong interaction threshold parameters, which were not considered
above (these are not yet published in the literature in a definite form).
A close analysis of the results obtained above shows, however, that
eventual uncertainties on the threshold or low energy parameters have
the tendancy to move the predicted value of the lifetime by remaining
within the band of uncertainty already considered. An illustration of
this phenomenon can be seen by considering the set II solution obtained
in the standard $\chi PT$ case \cite{bcegs} and for which one has in
particular $(a_0^0-a_0^2)=0.250$. (A recent analysis of the threshold
parameters in the standard $\chi PT$ case is presented in Ref. 
\cite{abt}.) One might for instance consider the
differences between predictions of set I and set II as uncertainties. One
then has two possibilities of proceeding. First, one can calculate directly
from set II the value of the lifetime, by repeating the calculations
done with set I. Second, one can seek from the curve of Fig. \rf{f1}
the value of the lifetime corresponding to $(a_0^0-a_0^2)=0.250$. For
the first method, we have found from the threshold parameters of set II
the following values of the low energy constants: $\alpha=1.010$,
$\beta=1.111$, $\lambda_1=-9.11\times 10^{-3}$,
$\lambda_2=11.15\times 10^{-3}$, $\lambda_3=-1.14\times 10^{-4}$,
$\lambda_4=-0.27\times 10^{-4}$. The predicted value of the lifetime is
$\tau=3.22$ fs. This value lies within the band of uncertainty of Fig.
\rf{f1}. The value one obtains for $(a_0^0-a_0^2)=0.250$ from Fig. \rf{f1},
corresponding to the initial data of set I, is
$\tau=(3.20\pm 0.07)$ fs, which underlines the fact that uncertainties
of the order of 3\% in the scattering length values have not moved the 
lifetime value outside the initial uncertainty band (and furthermore
have left it rather close to the central line). Therefore, one should not
enlarge further the initial uncertainty band of 2-2.5\%, which could be
considered as a conservative one.
\par
One thus arrives at the conclusion that the relationship
between the pionium lifetime and the combination $(a_0^0-a_0^2)$ of the
$S$-wave scattering lengths represented in Fig. \rf{f1}, together
with its uncertainty band, is practically independent of the particular
choice of the low energy constants which yield the value of
$(a_0^0-a_0^2)$. In this sense, the curve of Fig. \rf{f1} should be used
to determine from the experimental value of the pionium lifetime (with
uncertainties) the corresponding value of $(a_0^0-a_0^2)$ (with
corresponding uncertainties). The determination of the other thereshold
parameters or low energy constants, and in particular of $\alpha $ and of
the quark condensate parameter, requires a separate analysis with the aid
of additional constraints, coming for instance from sum rules and the
Roy equations.
\par
It is also evident that for a resulting value of $(a_0^0-a_0^2)$,
because of the other existing constraints, the value of $\alpha$, and
hence of the quark condensate parameter, could not be arbitrarily
varied. Independent of any detailed analysis which should determine the
precise value of $\alpha$, one still is allowed to derive from Fig. \rf{f1}
some qualitative conclusions which we summarize as follows. Values of
the lifetime close to 3 fs, lying above 2.9 fs, say, would confirm the
scheme of standard $\chi PT$. Values of the lifetime lying below 2.4 fs
remain outside the domain of predictions of standard $\chi PT$ and would
necessitate an alternative scheme of chiral symmetry breaking. Values of the
lifetime lying in the interval 2.4-2.9 fs, because of the possibly existing
uncertainties, would be more difficult to interpret and would require
a more refined analysis; they might also indicate, within the standard 
$\chi PT$ scheme, a strong dependence of the quark condensate on the light
quark flavor number, as a trace of a possible chiral phase transition
at higher values of the latter.
\par

\noindent
\textbf{Acknowledgments}:
I thank J. Gasser, L. Girlanda, M. Knecht, L. Nemenov, A. Rusetsky and J.
Stern for useful discussions.
\par

\end{document}